\documentclass[preprint, 3p, number]{elsarticle}

\usepackage{setspace}

\usepackage{amssymb}
\usepackage{textcomp}  
\usepackage{comment}
\usepackage{subfigure}
\usepackage{tablefootnote}

\usepackage{setspace}

\newcounter{daggerfootnotea}
\newcommand*{\daggerfootnotea}[1]{%
    \setcounter{daggerfootnotea}{\value{footnote}}%
    \renewcommand*{\thefootnote}{\fnsymbol{footnote}}%
    \footnote[2]{#1}\textsuperscript{,}%
    \setcounter{footnote}{\value{daggerfootnotea}}%
    \renewcommand*{\thefootnote}{\arabic{footnote}}%
    }
    
\newcounter{daggerfootnoteb}
\newcommand*{\daggerfootnoteb}[1]{%
    \setcounter{daggerfootnoteb}{\value{footnote}}%
    \renewcommand*{\thefootnote}{\fnsymbol{footnote}}%
    \footnote[3]{#1}\textsuperscript{,}%
    \setcounter{footnote}{\value{daggerfootnoteb}}%
    \renewcommand*{\thefootnote}{\arabic{footnote}}%
    }
    
\newcounter{daggerfootnotec}
\newcommand*{\daggerfootnotec}[1]{%
    \setcounter{daggerfootnotec}{\value{footnote}}%
    \renewcommand*{\thefootnote}{\fnsymbol{footnote}}%
    \footnote[4]{#1}\textsuperscript{,}%
    \setcounter{footnote}{\value{daggerfootnotec}}%
    \renewcommand*{\thefootnote}{\arabic{footnote}}%
    }

\usepackage{hyperref}
\usepackage[pdftex]{color}
\usepackage[font=footnotesize,labelfont=bf]{caption}

\graphicspath{{./}{Figures/}}

\hypersetup{
    colorlinks=true,
    linkcolor=blue,
    filecolor=magenta,      
    urlcolor=cyan,
    pdftitle={Overleaf Example},
    pdfpagemode=FullScreen,
    }

\usepackage{lineno}

\journal{Astroparticle Physics}

\begin{document}

\begin{frontmatter}



\title{KamLAND's search for correlated low-energy electron antineutrinos with astrophysical neutrinos from IceCube}


\author[tohoku]{S.~Abe}
\author[tohoku]{S.~Asami}
\author[tohoku]{M.~Eizuka}
\author[tohoku]{S.~Futagi}
\author[tohoku]{A.~Gando}
\author[tohoku]{Y.~Gando}
\author[tohoku]{T.~Gima}
\author[tohoku]{A.~Goto} 
\author[tohoku]{T.~Hachiya}
\author[tohoku]{K.~Hata} 

\author[tohoku]{K.~Hosokawa\daggerfootnotea{Present address: Kamioka Observatory, Institute for Cosmic-Ray Research, The University of Tokyo, Hida, Gifu
506-1205, Japan}} 
\author[tohoku]{K.~Ichimura} 
\author[tohoku]{S.~Ieki} 
\author[tohoku]{H.~Ikeda}
\author[tohoku,ipmu]{K.~Inoue}

\author[tohoku]{K.~Ishidoshiro}
\author[tohoku]{Y.~Kamei}
\author[tohoku]{N.~Kawada} 
\author[tohoku,ipmu]{Y.~Kishimoto} 
\author[tohoku]{T.~Kinoshita} 
\author[tohoku,ipmu]{M.~Koga}
\author[tohoku]{M.~Kurasawa}

\author[tohoku]{N.~Maemura}
\author[tohoku]{T.~Mitsui}
\author[tohoku]{H.~Miyake}
\author[tohoku]{T.~Nakahata}
\author[tohoku]{K.~Nakamura}
\author[tohoku]{K.~Nakamura}
\author[tohoku]{R.~Nakamura}
\author[tohoku,gppu]{H.~Ozaki}
\author[tohoku]{T.~Sakai} 
\author[tohoku]{H.~Sambonsugi}
\author[tohoku]{I.~Shimizu}
\author[tohoku]{J.~Shirai}
\author[tohoku]{K.~Shiraishi}
\author[tohoku]{A.~Suzuki}
\author[tohoku]{Y.~Suzuki}
\author[tohoku]{A.~Takeuchi\daggerfootnoteb{Present address: Department of Physics, Faculty of Science, the University of Tokyo, Tokyo 113-0033, Japan}}
\author[tohoku]{K.~Tamae}
\author[tohoku]{H.~Watanabe}
\author[tohoku]{Y.~Yoshida} 

\author[fris]{S.~Obara\daggerfootnotec{Present address: National Institutes for Quantum and Radiological Science and Technology (QST), Sendai 980-8579,Japan}}

\author[tohokuRigaku]{A.~K.~Ichikawa}
\author[osakarcnp]{S.~Yoshida}
\author[osaka]{S.~Umehara}

\author[tokushima]{K.~Fushimi}
\author[tokushima]{K.~Kotera}
\author[tokushima]{Y.~Urano}

\author[lbl,ipmu]{B.~E.~Berger}
\author[lbl,ipmu]{B.~K.~Fujikawa}

\author[hawaii]{J.~G.~Learned}
\author[hawaii]{J.~Maricic}

\author[mituniv]{S.~N.~Axani}
\author[mituniv]{J.~Smolsky}
\author[mituniv]{J.~Lertprasertpong}
\author[mituniv]{L.~A.~Winslow}
\author[mituniv]{Z.~Fu}
\author[mituniv]{J.~Ouellet}

\author[tennessee,ipmu]{Y.~Efremenko}

\author[tunl,chapehill]{H.~J.~Karwowski}
\author[tunl,northcarolina]{D.~M.~Markoff}
\author[tunl,duke,ipmu]{W.~Tornow}
\author[chapehill]{A.~Li}

\author[seattle,ipmu]{J.~A.~Detwiler}
\author[seattle,ipmu]{S.~Enomoto}

\author[nikhef,ipmu]{M.~P.~Decowski}

\author[bu]{C.~Grant}
\author[bu]{H.~Song}

\author[virginia]{T.~O'Donnell}
\author[virginia]{S.~Dell'Oro}


\affiliation[tohoku]{organization={Research Center for Neutrino Science},
            addressline={Tohoku University}, 
            city={Sendai},
            postcode={980-8578}, 
            country={Japan}}

\affiliation[fris]{organization={Frontier Research Institute for Interdisciplinary Sciences},
            addressline={Tohoku University}, 
            city={Sendai},
            postcode={980-8578}, 
            country={Japan}}
            
\affiliation[gppu]{organization={Graduate Program on Physics for the Universe},
            addressline={Tohoku University}, 
            city={Sendai},
            postcode={980-8578}, 
            country={Japan}}
            
\affiliation[tohokuRigaku]{organization={Department of Physics},
            addressline={Tohoku University}, 
            city={Sendai},
            postcode={980-8578}, 
            country={Japan}}
            
\affiliation[ipmu]{organization={Institute for the Physics and Mathematics  of the Universe},
            addressline={The University of Tokyo}, 
            city={Kashiwa},
            postcode={277-8568}, 
            country={Japan}}

\affiliation[osakarcnp]{organization={Graduate School of Science},
            addressline={Osaka University}, 
            city={Toyonaka},
            postcode={560-0043}, 
            country={Japan}}

\affiliation[osaka]{organization={Research Center for Nuclear Physics (RCNP)},
            addressline={Osaka University, Ibaraki}, 
            city={Osaka},
            postcode={567-0047}, 
            country={Japan}}

\affiliation[tokushima]{organization={Graduate School of Advanced Technology and Science},
            addressline={Tokushima University}, 
            city={Tokushima},
            postcode={770-8506}, 
            country={Japan}}
            


\affiliation[lbl]{organization={Nuclear Science Division},
            addressline={Lawrence Berkeley National Laboratory}, 
            city={Berkeley},
            state = {CA},
            postcode={94720}, 
            country={USA}}

\affiliation[hawaii]{organization={Department of Physics and Astronomy},
            addressline={University of Hawaii at Manoa}, 
            city={Honolulu},
            state = {HI},
            postcode={96822}, 
            country={USA}}
            
\affiliation[mituniv]{organization={Massachusetts Institute of Technology},
            city={Cambridge},
            postcode={02139}, 
            state={MA},
            country={USA}}
            
\affiliation[bu]{organization={Boston University},
            city={Boston},
            state = {MA},
            postcode={02215}, 
            country={USA}}
            
\affiliation[tennessee]{organization={Department of Physics and Astronomy},
            addressline={University of Tennessee}, 
            city={Knoxville},
            state = {TN},
            postcode={37996}, 
            country={USA}}

\affiliation[tunl]{organization={Triangle Universities Nuclear Laboratory},
            city={Durham},
            state = {NC},
            postcode={27708}, 
            country={USA}}

\affiliation[chapehill]{organization={The University of North Carolina at Chapel Hill},
            city={Chapel Hill},
            state = {NC},
            postcode={27599}, 
            country={USA}}

\affiliation[northcarolina]{organization={North Carolina Central University},
            city={Durham},
            state = {NC},
            postcode={27701}, 
            country={USA}}

\newcommand{\duke}{\affiliation{Physics Department at Duke University, Durham, NC 27705, USA}}
\affiliation[duke]{organization={Physics Department at Duke University},
            city={Durham},
            state = {NC},
            postcode={27705}, 
            country={USA}}

\affiliation[seattle]{organization={Center for Experimental Nuclear Physics and Astrophysics},
            addressline={University of Washington}, 
            city={Seattle},
            state = {WA},
            postcode={98195}, 
            country={USA}}

\affiliation[nikhef]{organization={Nikhef and the University of Amsterdam},
            addressline={Science Park}, 
            city={Amsterdam},
            country={the Netherlands}}

\affiliation[virginia]{organization={Center for Neutrino Physics},
            addressline={ Virginia Polytechnic Institute and State University}, 
            city={Blacksburg},
            state = {VA},
            postcode={24061}, 
            country={USA}}

\begin{abstract}
We report the results of a search for MeV-scale astrophysical neutrinos in KamLAND presented as an excess in the number of coincident neutrino interactions associated with the publicly available high-energy neutrino datasets from the IceCube Neutrino Observatory. We find no statistically significant excess in the number of observed low-energy electron antineutrinos in KamLAND, given a coincidence time window of $\pm$500\,s, $\pm$1,000\,s, $\pm$3,600\,s, and $\pm$10,000\,s around each of the IceCube neutrinos. We use this observation to present limits from 1.8\,MeV to 100\,MeV on the electron antineutrino fluence, assuming a mono-energetic flux. We then compare the results to several astrophysical measurements performed by IceCube and place a limit at the 90\% confidence level on the electron antineutrino isotropic thermal luminosity from the TXS\,0506+056 blazar. 
\end{abstract}




\end{frontmatter}

\section{Introduction} \label{sec::intro}

Astrophysical neutrinos are capable of delivering unprecedented information about the most cataclysmic events in the Universe. Since neutrinos only interact weakly with matter and are not deflected by galactic/intergalactic magnetic fields, they point back to their production source and carry information relating to the \textit{in situ} physical conditions of some of nature's most extreme environments. 
In 2013, the IceCube Neutrino Observatory discovered the diffuse astrophysical neutrino flux~\cite{icecube_astro,aartsen2015atmospheric,aartsen2020characteristics}. This discovery introduced astrophysical neutrinos into the toolkit for extragalactic multimessenger astronomy and was soon followed by the observation of a high-energy muon neutrino on 22 September 2017 (IceCube-170922A) in coincidence with, and in the direction of, a flaring gamma-ray blazar (TXS 0506+056). The significance of this observation is estimated to be at the 3$\sigma$ level~\cite{IC2018}.  When analyzing the antecedent neutrino flux originating from the direction of TXS 0506+056, in conjunction with data collected by Large Area Telescope (LAT) on the Fermi Gamma-ray Space Telescope~\cite{tanaka2017fermi} and the Major Atmospheric Gamma Imaging Cherenkov (MAGIC) telescopes~\cite{aleksic2012performance,mirzoyan2017first}, the  significance of the time-dependent excess was observed at the $3.5\sigma$ level~\cite{IC2018}, statistically independent of the 2017 flaring episode. 
These observations provided the most compelling evidence supporting the long-suspected theory that jetted active galactic nuclei (AGN) contribute to the highest energy ($\gtrsim$100\,TeV) extragalactic neutrino background\,\cite{halzen2005high, adrian2012search,kadler2016coincidence}. 
The most recent studies have reported a 2.9$\sigma$ neutrino excess at the coordinates of NGC 1068~\cite{aartsen2020time}, a radio quiet AGN; a 2.6$\sigma$ excess compatible with neutrino production in the cores of AGNs~\cite{abbasi2021search}; the likely association of a 0.2\,PeV neutrino (IC191001A) with a tidal disruption event, AT2019dsg~\cite{stein2021tidal}; an optical outburst from an AGN, AT2019fdr, coincident with a 0.08\,PeV neutrino, IC200530A~\cite{reusch2021candidate}; and a coincidence PeV-scale neutrino with the transient emission from a super-massive black hole, AT2019aalc~\cite{van2021establishing}.




An international effort has recently emerged to develop the next generation of high-energy astrophysical neutrino observatories: KM3NeT-ORCA and KM3NeT-ARCA~\cite{adrian2016letter} off the coast of France and Sicily; Baikal-GVD~\cite{baikal1808baikal} in Siberia; IceCube-Gen2~\cite{aartsen2021icecube} in Antarctica; and the Pacific Ocean Neutrino Experiment (P-ONE)~\cite{agostini2020pacific} off the west coast of Vancouver Island in Canada. 

While the sources of the high-energy astrophysical neutrinos may soon be determined,
the correlation between MeV and TeV-to-PeV neutrino production remains largely unconstrained. MeV neutrinos have been observed from the core-collapse supernovae (CCSNe) SN 1987A~\cite{hirata1988observation,schramm1987neutrinos,lattimer1989analysis}, and some CCSNe are good
candidates for the production of high-energy neutrinos through the
interaction of ejecta with circumstellar material~\cite{murase2018new} and through choked jets~\cite{razzaque2005high} (as suggested by a SN 1987A followup analysis presented in Ref.~\cite{oyama2022evidence}). Many astrophysical objects exhibit spectra characteristic of an ultra-relativistic collimated jet formation (such as binary mergers and gamma-ray bursts) that can give rise to a high-energy secondary neutrino beam. These objects may also produce MeV neutrinos in the resulting accretion disk, disk corona, molecular wind/outflow from these objects, and/or hot dense remnant~\cite{rosswog2003high, setiawan2006three}. In any case, the models involve significant astrophysical uncertainties~\cite{meszaros2017astrophysical,murase2017active,learned2000high}, and motivate experimental searches that branch the extreme energy ranges. The Super-Kamiokande experiment has recently published the null observation of a 225\,kiloton-years search for the GeV astrophysical neutrino counterpart~\cite{abe2017search}, and the null observation of a 22-year GeV to several TeV search specifically looking for a neutrino excess in the direction of TXS\,0506+056~\cite{hagiwara2019search}.  


In this analysis, we present a search for time-correlated MeV electron antineutrino ($\bar{\nu}_e$) events in the Kamioka Liquid Scintillator Antineutrino Detector (KamLAND) associated with several publicly available high-energy neutrino datasets from IceCube. We include data spanning from the earliest IceCube data on 9 October 2010, through 12 June 2021. 
We use KamLAND electron antineutrinos with energies ranging from 1.8\,MeV$<E_{\bar{\nu}_e}<$100\,MeV, thus making this the lowest energy astrophysical coincident neutrino search associated with IceCube to date. Observing the low-energy counterpart of the astrophysical neutrino flux would not only help with source identification but also with understanding the astrophysical neutrino production mechanism. 

\section{KamLAND detector} \label{sec:detector} 
\begin{figure}[t]
\centering
\includegraphics[width=0.65\linewidth]{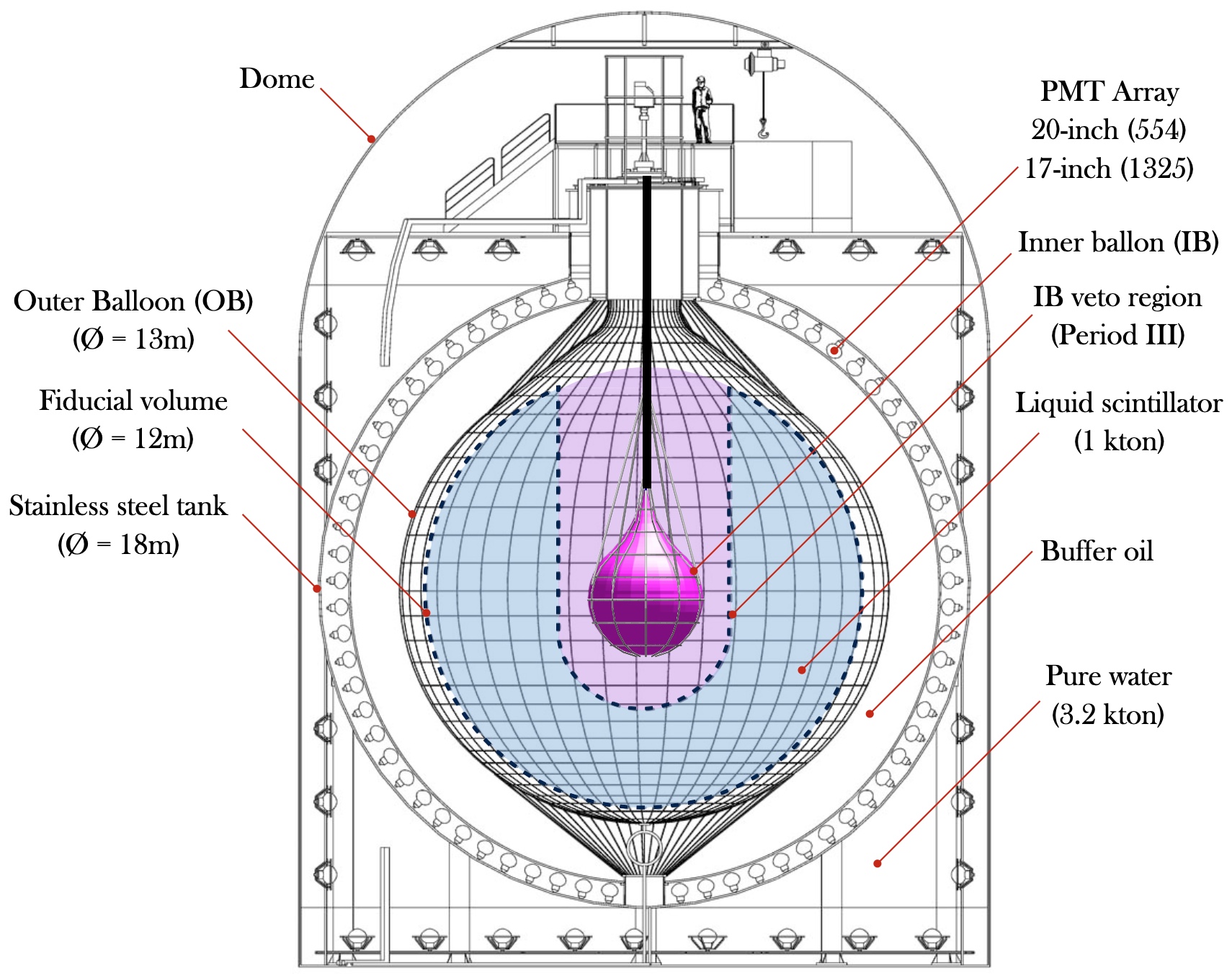}
\caption{A schematic of the KamLAND detector. The fiducial volume for this analysis is shown in blue. The geometric cut for Periods II \& IV is shown in purple. Image adapted from Ref.~\cite{abe2022searchgrb}.}
\label{fig::kam}
\end{figure}

KamLAND is a kiloton-class liquid scintillator neutrino detector situated 1\,km below the surface of Mt.\,Ikenoyama  in Kamioka, Japan. KamLAND is separated into an inner and outer detector, demarcated by an 18\,m spherical stainless steel tank (illustrated in Fig.~\ref{fig::kam}). The inner detector was primarily optimized for MeV electron antineutrino ($\bar{\nu}_e$) detection~\cite{gando2013reactor, araki2005experimental,abe2008precision}. It contains a 13\,m diameter transparent balloon holding approximately 1,200\,m$^3$ of ultrapure liquid scintillator. Scintillation light is observed by 1,325 17-inch photomultipliers (PMT) and 554 20-inch PMTs suspended in a non-scintillating buffer-oil solution. The PMTs point radially inwards into the inner detector providing approximately 34\% photocathode coverage. This analysis considers the fiducial volume of the detector to be the innermost 12\,m diameter region of the inner detector. The outer detector contains 3.2\,kton of pure water that acts as a cosmic-ray muon veto and passive shield against radioactivity originating in the surrounding material. The full KamLAND detector is synchronized to an external Global Positioning System receiver placed outside the entrance to the Kamioka mine, providing a global time uncertainty of less than 100\,$\mu s$. 

From August 2011 up to October 2015, a transparent teardrop-shaped, 3.0\,m-diameter inner balloon containing approximately 350\,kg xenon with an enriched concentration of $^{136}$Xe was installed into the central region of the inner detector as part of the KamLAND-Zen\,400 experiment~\cite{gando2016search}. This was replaced with a larger inner balloon containing approximately 700\,kg of enriched xenon in April 2018 for the KamLAND-Zen\,800 experiment~\cite{zen2022first,gando2020first,gando2021nylon}.
\section{Electron antineutrino selection in KamLAND and background estimation} \label{sec:antiv}
We consider the KamLAND events triggered by the inverse beta decay (IBD) interaction, where an $\bar{\nu}_e$ interacts weakly with a proton producing a positron and neutron in the final state ($\bar{\nu}_e +p \rightarrow e^{+}+n$).  In this interaction, the positron quickly deposits its kinetic energy into the scintillator and annihilates with an electron near the primary interaction vertex, creating a \textit{prompt} signal. The neutron thermalizes via elastic scatters before capturing on a proton producing a deuteron and a 2.2\,MeV gamma ray. A small fraction (0.5\%\,\cite{abe2010production}) of the neutron captures  occur on a carbon-12 nucleus, producing a 4.9\,MeV gamma ray. The neutron has a mean capture time of 207.5\,$\pm$\,2.8\,$\mu s$\,\cite{abe2010production} and the emission of the gamma ray is observed as a \textit{delayed} signal. The time and space correlation between the successive delayed coincidence (DC) signals allows for efficient background suppression, enabling a high-efficiency detection of IBD interactions. Specifically, we require the delayed interaction vertex to be located within 200\,cm of the prompt signal, and the time difference between the delayed and prompt signal to be less than 1000\,$\mu$s.  The interaction vertex resolution in KamLAND can be approximated as $12$\,cm\,$/\sqrt{E_{\bar{\nu}_e}(\mathrm{MeV})}$~\cite{gando2013reactor}. Since the  angular distribution of the positron is nearly isotropic and the scintillation light emission is also isotropic, KamLAND is unable to reconstruct the direction of the incident $\bar{\nu}_e$.

The IBD interaction has a kinematic low-energy neutrino threshold of $E_{\bar{\nu}_e} = 1.806$\,MeV. We can relate $E_{\bar{\nu}_e}$ to the measured prompt energy, $E_{p}$, via $E_{\bar{\nu}_e} \approx E_{p} + 0.78$\,MeV\,+\,$T_n$, where $T_n$ is the neutron kinetic energy. The thermalization of the neutron is also contained in the prompt signal, however it is sufficiently quenched that it can be ignored in the energy range of this analysis. The delayed energy is selected to be between 1.8\,MeV\,$\leq E_d \leq 2.6$\,MeV or 4.4\,MeV\,$\leq E_d \leq 5.6$\,MeV and the prompt energy is selected to be between 0.9\,MeV\,$\leq$\,$E_{p}$\,$\leq$\,93\,MeV. The IBD selection criteria also include a likelihood-based series of cuts capable of further reducing the background, described in Ref.\,~\cite{gando2013reactor}. 
We approximate the energy resolution of KamLAND as $6.4 \%/\sqrt{E_{\bar{\nu}_e} (\mathrm{MeV})}$~\cite{gando2013reactor}. 

The KamLAND IBD data used in this analysis is separated into four time periods. Period I refers to the data collected prior to the installation of the KamLAND-Zen\,400 inner balloon on 12 October 2011. This period includes an additional cut on the prompt energy, such that events with $E_{\bar{\nu}_e} < 8.3$\,MeV are rejected in order to reduce the reactor neutrino background. Since all Japanese nuclear reactors were shut down in March 2011, due to the Great East Japan Earthquake, subsequent periods do not include this cut. Period\,II refers to the data taking period in which the KamLAND-Zen\,400 inner balloon was installed in the detector (12 October 2011 to 24 December 2015). Here, we include an additional geometric cut on the delayed event around the inner balloon and support structure to reduce backgrounds introduced by the additional material. The cut removes events in which the delayed signal occurs within the central spherical 2.5\,m radius region, extending to the top of the detector in a 2.5\,m radius cylinder. We account for the reduction in fiducial volume due to this cut in the detection efficiency rather than the number of target nuclei. Period\,III spans the time between the extraction of the KamLAND-Zen\,400 inner balloon (24 December 2015) and Jan. 2 2019, the start of KamLAND-Zen\,800. During this period, we do not include the 2.5\,m radius geometric cut. The final period, Period IV, includes all data from the introduction of the KamLAND-Zen\,800 inner balloon (16 April 2018) onwards. Here, we also include the same geometric cut described for Period II. The acquisition time and livetime, along with the livetime efficiency, $\epsilon_{\mathrm{live}}$, for each period are shown in the leftmost columns of Table~\ref{table::PeriodData}. 


KamLAND data is recorded in runs, which tend to span approximately one day. We impose quality checks on each run to ensure that the detector was in suitable operating condition and not taking data during calibration. The final IBD event sample was found to contain 341 IBD events. We consider all IBD interactions in the background calculation. 
Table~\ref{table::PeriodData} shows the number of IBD events observed in each period during the detector livetime. 

In Period I, the IBD sample is dominated by atmospheric neutrinos. There is also a significant contribution arising from long-lived spallation products and fast neutrons from cosmic-ray muons~\cite{gando2012search}. While these are also the dominant background above E$_{\bar{\nu}e} > 8.3$\,MeV in Period II\,-\,IV, below this energy neutrinos originating from the small number of operational nuclear reactor power plants dominate the IBD sample. At energies below E$_{\bar{\nu}e} \sim$ 3.4\,MeV, there is also a contribution from radiogenic neutrinos originating from the decay of $^{232}$Th and $^{238}$U within the Earth. The IBD sample also includes a small number of accidental DC signals induced by the decay of radioactive impurities, additional radioactive products from cosmic-ray muon spallation~\cite{abe2010production,abe2022limits}, and the alpha-induced $^{13}$C($\alpha$,n)$^{16}$O reaction in the liquid scintillator.

\begin{table}[h]
\setlength\tabcolsep{2pt} 
\caption{The period-specific data used in this analysis, assuming a time window of $\pm500$\,s.  
The number of IBD events and IceCube candidate events found in the livetime of each period is shown in the forth and fifth column.
Given the total number of expected background events in each period along with the null observation of a coincident event, the last column shows the corresponding Feldman-Cousins 90\% C.L. upper limit on the number of observed signal neutrinos. }
\label{table::PeriodData}
\normalsize
\centering
\begin{tabular}{ c c c c c c c c c c }
\hline
\hline
{} & Realtime & Livetime & $\epsilon_{\mathrm{live}}$ &  IBD & IC Events & Total Window & Total Bkg. & Total Sig. &  N$_{90}$ \\ 
{} & [days] & [days] & [\%] & Counts  & Counts & [h]& Counts & Counts & Counts\\[0.5ex]
\hline
    Period I   & 393.4 & 244.7& 62.2 & 3 & 4        & 1.1 & 5.7$\times$10$^{-4}$ & 0 & 2.43 \\ [0.5ex]
    Period II  & 1471.9 & 1370.4 & 93.1 & 148 & 22  & 6.1 & 2.7$\times$10$^{-2}$ & 0 & 2.40 \\ [0.5ex]
    Period III & 561.0 & 482.0 & 85.9 & 58 & 14     & 3.9 & 2.0$\times$10$^{-2}$ & 0 & 2.41 \\ [0.5ex]
    Period IV  & 1152.7 & 1094.0 & 94.9 & 132 & 62  & 17.2& 8.7$\times$10$^{-2}$ & 0 & 2.34 \\ [0.5ex]
    \hline
    Total      & 3579.0 & 3191.1 & 89.2 & 341 & 102 & 28.3 & $13.4\times$10$^{-2}$ & 0 & 2.30   \\ [0.5ex]
    \hline
\end{tabular}
\end{table}

\section{IceCube high-energy neutrino sample} \label{sec:IC}
Historical IceCube neutrino events that met the criteria of the Extremely High Energy (EHE) and High Energy Starting Event (HESE) IceCube filters were taken from a publicly available  data release\footnote{\url{https://icecube.wisc.edu/data-releases/2018/07/}}. The first of these events was on 9 October 2010. The historical EHE filter contained 24 track-like (muon neutrino) events originating from the Northern Hemisphere with neutrino energies greater than several hundred\,TeV. The historical HESE filter contained 16 track-like events whose interaction vertices originated within the instrumented volume of IceCube, five of which have since been retracted due to poor angular reconstruction. Since the HESE events start within the instrumented volume, this filter is also capable of accepting events that originate from the Southern Hemisphere. These two filters were subsequently added to the realtime Astrophysical Multimessenger Observatory Network (AMON) in early 2016.  Depending on the signal, AMON distributes alerts in realtime for follow-up observations via the Gamma-ray Coordinates Network (GCN). In total, there have been nine EHE\footnote{\url{https://gcn.gsfc.nasa.gov/amon_ehe_events.html}} and eighteen HESE\footnote{\url{https://gcn.gsfc.nasa.gov/amon_hese_events.html}} realtime alerts issued. A single event was reported on both the AMON EHE and HESE system.

The EHE and HESE realtime alerts continued until 2019 May, when a new classification scheme referred to as ``Gold" and ``Bronze"\footnote{\url{https://gcn.gsfc.nasa.gov/amon_icecube_gold_bronze_events.html}} was introduced. The set of Gold events is expected to have an astrophysical purity greater than 50\%, whereas the purity of the Bronze sample is expected to be greater than  30\%. Since this change, there have been 22 Gold and 31 Bronze alerts issued prior to 12 June 2021, the end date of this analysis. A separate realtime alert classification also exists for the high-energy electron (anti)neutrino and neutral current events, known as the Cascade filter.\footnote{\url{https://gcn.gsfc.nasa.gov/amon_icecube_cascade_events.html}} This filter contains three events, all of which were issued in 2021, and is expected to have an astrophysical purity $>$\,85\%.

The IceCube-170922A event, discussed in Sec.~\ref{sec::intro}, is assumed to have originated from the flaring gamma-ray blazar, TXS 0506+056. The redshift was measured with Gran Telescopio Canarias and found to have a redshift of $z = 0.3365 \pm 0.0010$, corresponding to a luminosity distance of $d = 1.75$\,Gpc~\cite{Paiano_2018}. The IceCube-170922A event triggered the EHE filter and had a reconstructed energy of approximately 290\,TeV, with a 90\%
confidence level (C.L.) lower limit of 183\,TeV. It was reported on AMON with a follow-up campaign through the GCN from electromagnetic observatories.

Finally, an event was recently found to have a reconstructed visible energy of 6.05\,$\pm$\,0.72 PeV, consistent with a $\bar{\nu}_e$ interacting via the Glashow resonance~\cite{glashow}. The evidence of the Glashow resonance indicates the presence of $\bar{\nu}_e$s in the astrophysical flux. This event had its own data-release.\footnote{\url{https://icecube.wisc.edu/data-releases/2021/03/icecube-data-for-the-first-glashow-resonance-candidate/}} 

Of the 118 IceCube events described above, 102 arrived during the livetimes considered in this analysis. 
\begin{figure}[h]
\centering
\includegraphics[width=0.48\linewidth]{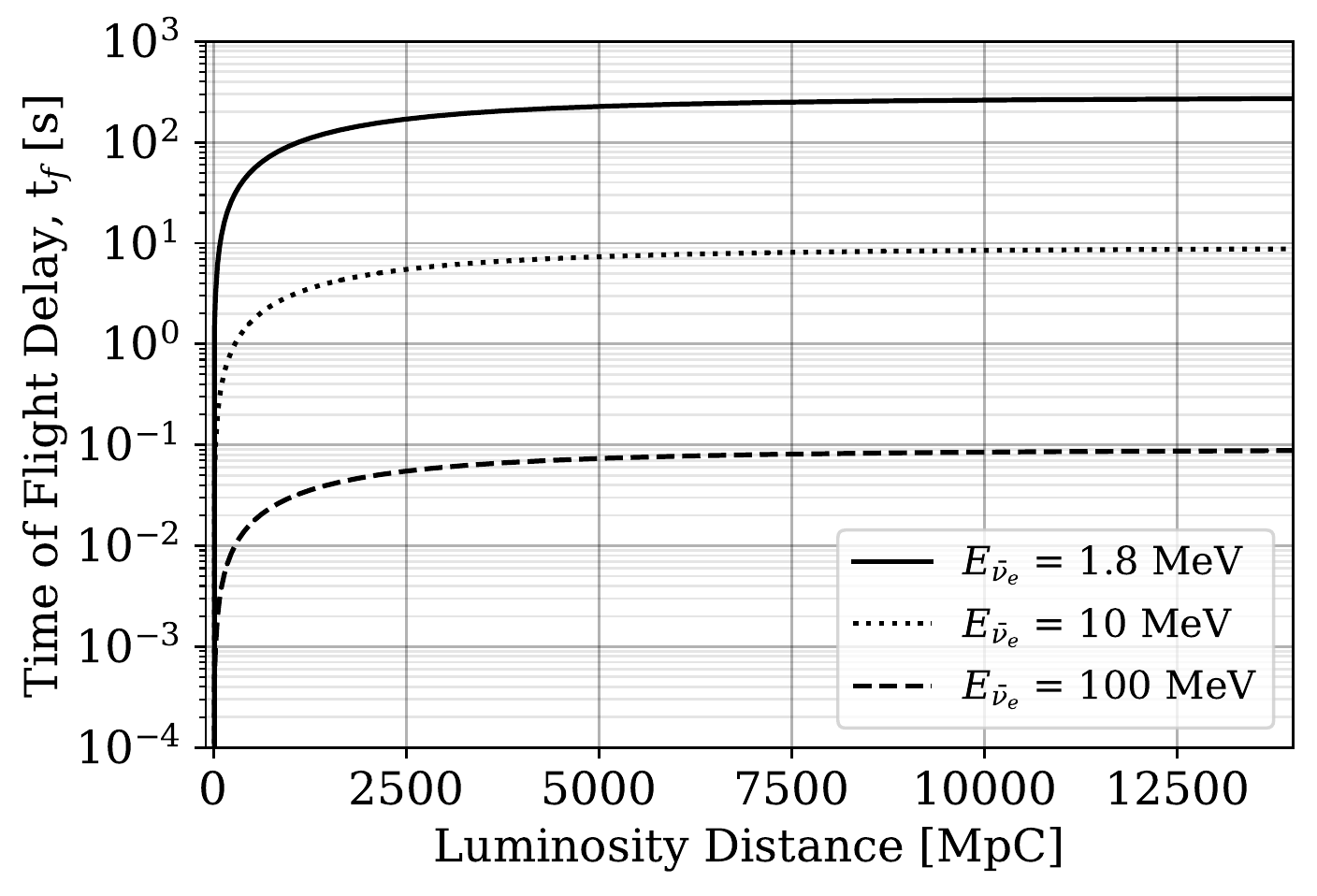}
\caption{A conservative estimate of the time-of-flight delay of a neutrino with an energy of 1.8\,MeV (solid), 10\,MeV (dotted), and 100\,MeV (dashed), relative to the speed of light. The time-of-flight delay of a 1.8\,MeV neutrino originating at the location of TXS 0506+056 is determined to be less than 138\,s.}
\label{fig::tof}
\end{figure}
\section{Correlated event search} \label{sec:candidates}
With no conclusive single source of astrophysical neutrinos or concrete theoretical production mechanism to link the high-energy to the low-energy neutrino flux, we attempt to reduce model-dependency by selecting various time windows to search for an excess number of coincident events. We define a coincidence search time window around each of the high-energy neutrinos in IceCube, $t_{HE}$, such that:
\begin{equation}\label{eq::window}
t_{HE} - t_{p} < t_{IBD} < t_{HE} +t_{p},
\end{equation} 
where $t_{IBD}$ is the IBD prompt signal global timestamp and $t_{p}$ represents a predefined window size that is sufficiently large to cover the low-energy neutrino time-of-flight delay and model dependency. 

The time-of-flight delay relative to the speed of light for a neutrino with redshift $z$, mass $m_\nu$, and energy $E_\nu$, can be calculated through~\cite{li2005testing}:
\begin{equation} 
t_f = \frac{1}{2 H_0} \frac{m_\nu ^2}{E_{\nu}^2} \int_{0}^{z}\frac{dz^\prime}{(1+z^\prime)^2 \sqrt{\Omega_{\Lambda}+\Omega_M (1+z^\prime)^3}},
\label{eq:tof}
\end{equation}
where we assume base-$\Lambda$CDM cosmology parameters from Ref.~\cite{aghanim2020planck}: Hubble constant $H_0$\,=\,67.4\,km\,s$^{-1}$\,Mpc$^{-1}$; matter density parameter $\Omega_M = 0.315$; and dark matter density parameter $\Omega_\Lambda = 0.685$.  The mass of the electron antineutrino is conservatively set to  m$_\nu$ = 0.087\,eV, that is, the approximate 90\% C.L. upper limit on the most massive neutrino mass eigenstate,  given a normal mass hierarchy with mass squared splittings from Ref.~\cite{esteban2020fate}, and using the sum of the neutrino masses to be $\sum $m$_\nu$ = 0.12\,eV~\cite{aghanim2020planck}. The time-of-flight delay, calculated from Eq.\,\ref{eq:tof}, is shown in Fig.\,\ref{fig::tof} for several neutrino energies. We find that even for the most distant source, the time of flight for a 1.8\,MeV neutrino is less than 270\,s. 

Given the maximum expected time-of-flight delay, we choose to define the minimal time window used in this analysis to be (a) $t_{p}=$ \,500\,s. This is a commonly adopted time window range used for similar analyses that search for correlated neutrinos with gravitational waves\,\cite{asakura2015study,abe2021search,Abe_2021} and gamma-ray bursts (GRB)~\cite{abe2022searchgrb,baret2011bounding,fukuda2002search}. Since there are some cosmological events that span more than 1000\,s, such as the longest duration GRBs, we also use several more conservative time windows of $t_{p} =$\,\,1000\,s (b),  \,3600\,s (c), and  \,10000\,s (d). This is also a common strategy for coincident searches with large uncertainties on the progenitor model~\cite{fukuda2002search}. If any of the time windows extend into a new KamLAND run or over a short period of detector deadtime, we ensure that at least 90\% of the total time window is covered by KamLAND livetime. 


\begin{figure}[h]
\centering
\begin{subfigure}
  \centering
  \includegraphics[width=0.244\linewidth]{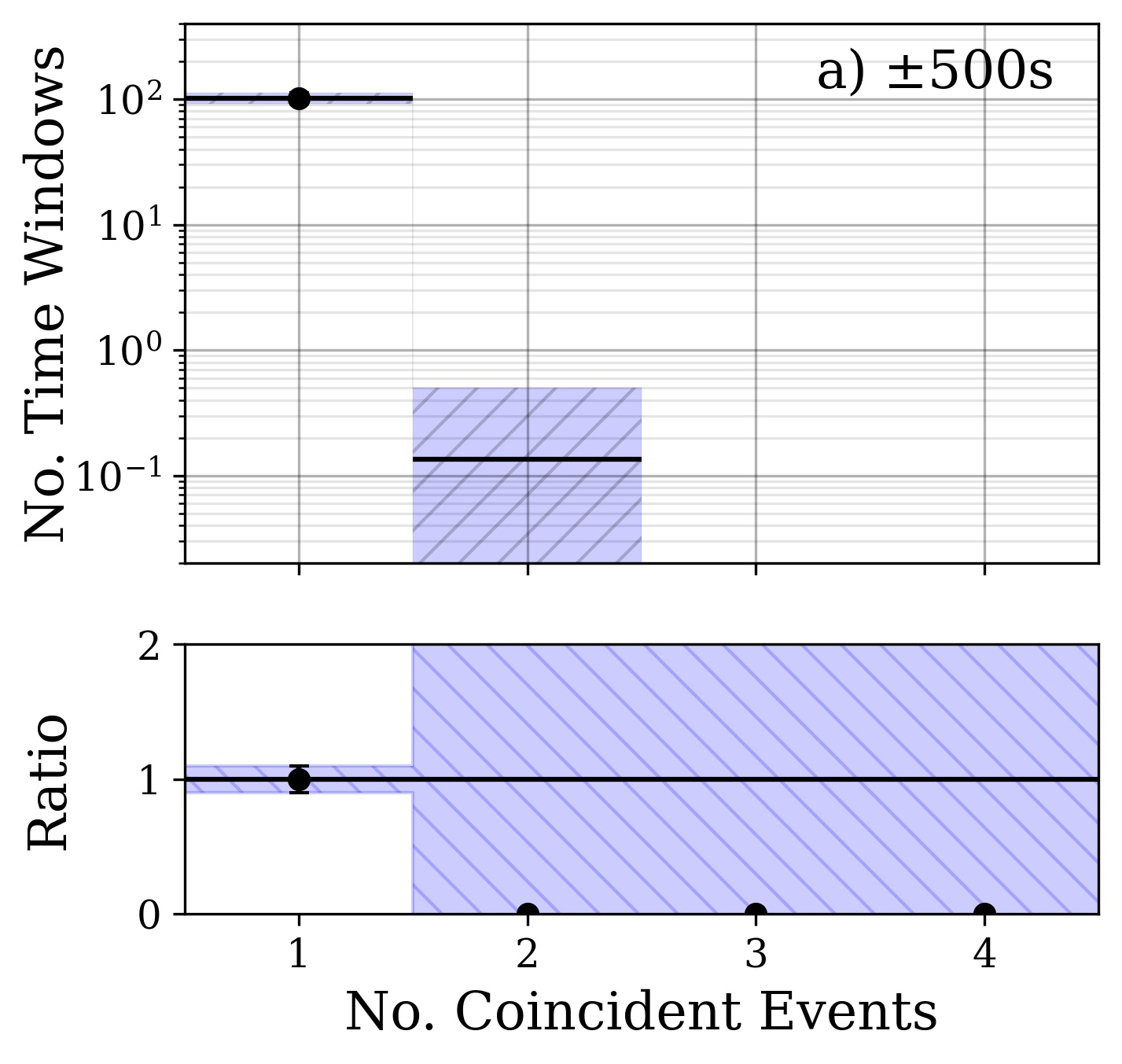}
\end{subfigure}
\begin{subfigure}
  \centering
  \includegraphics[width=.235\linewidth]{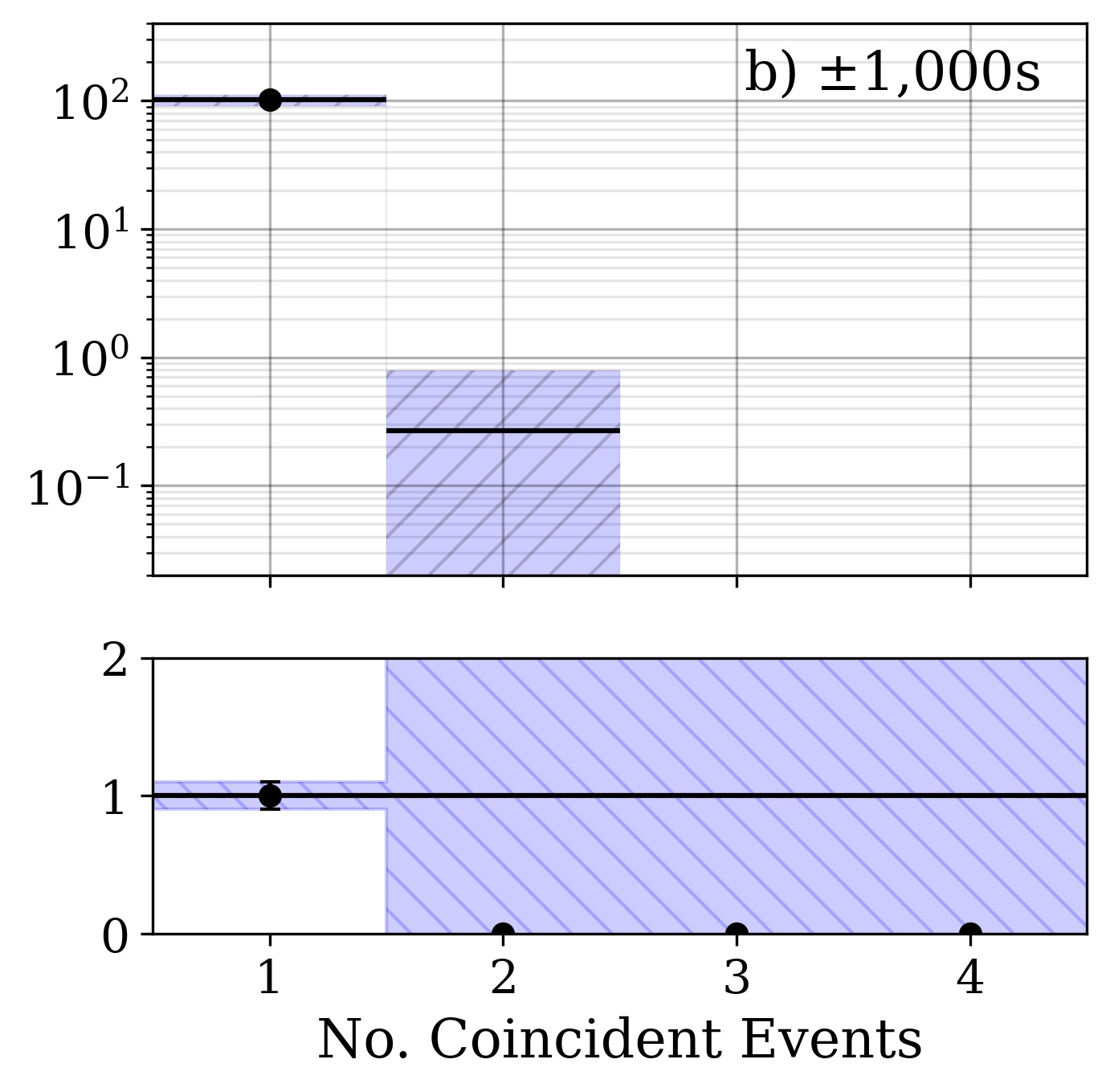}
\end{subfigure}
\begin{subfigure}
  \centering
  \includegraphics[width=.235\linewidth]{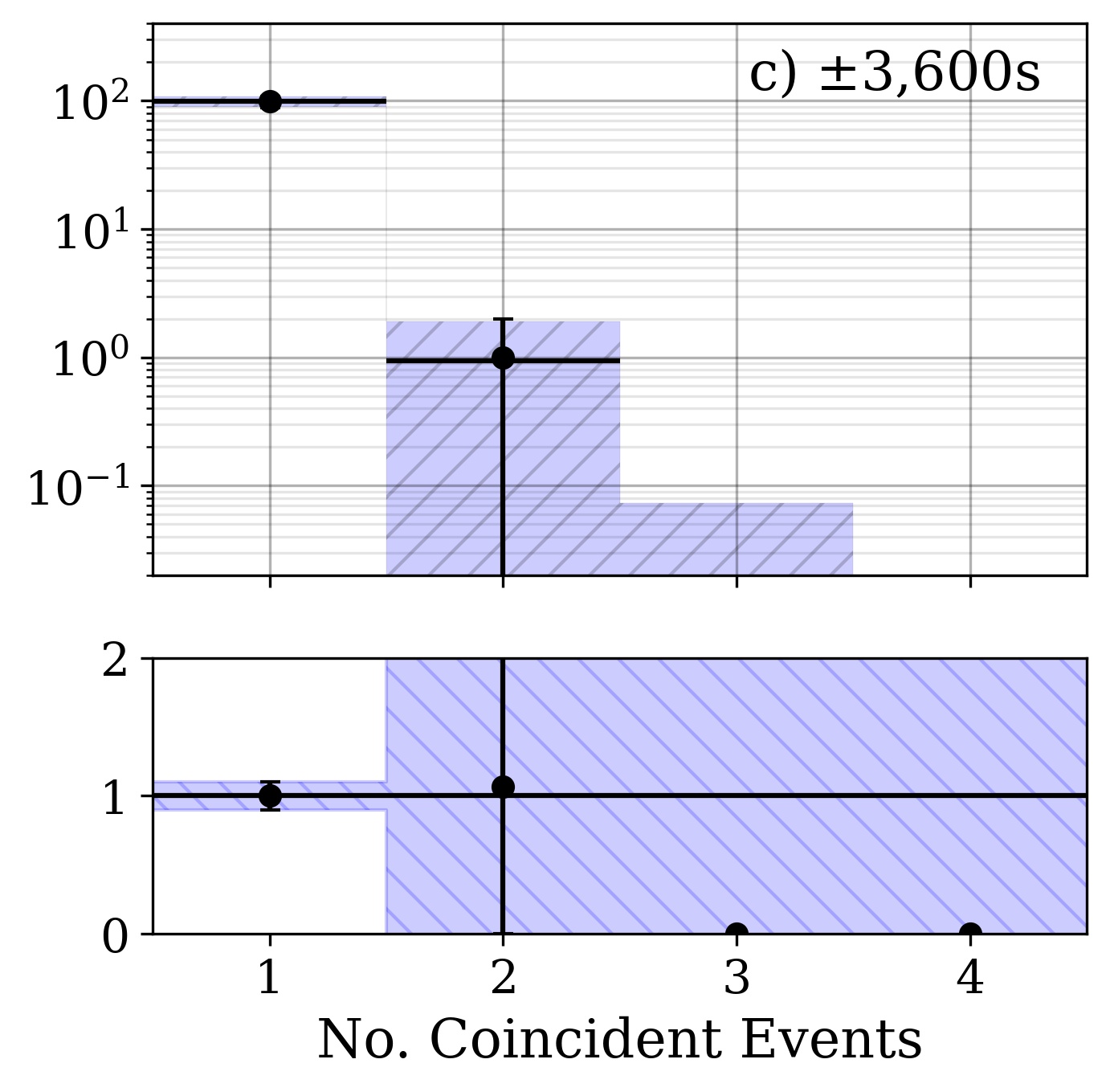}
\end{subfigure}
\begin{subfigure}
  \centering
  \includegraphics[width=.235\linewidth]{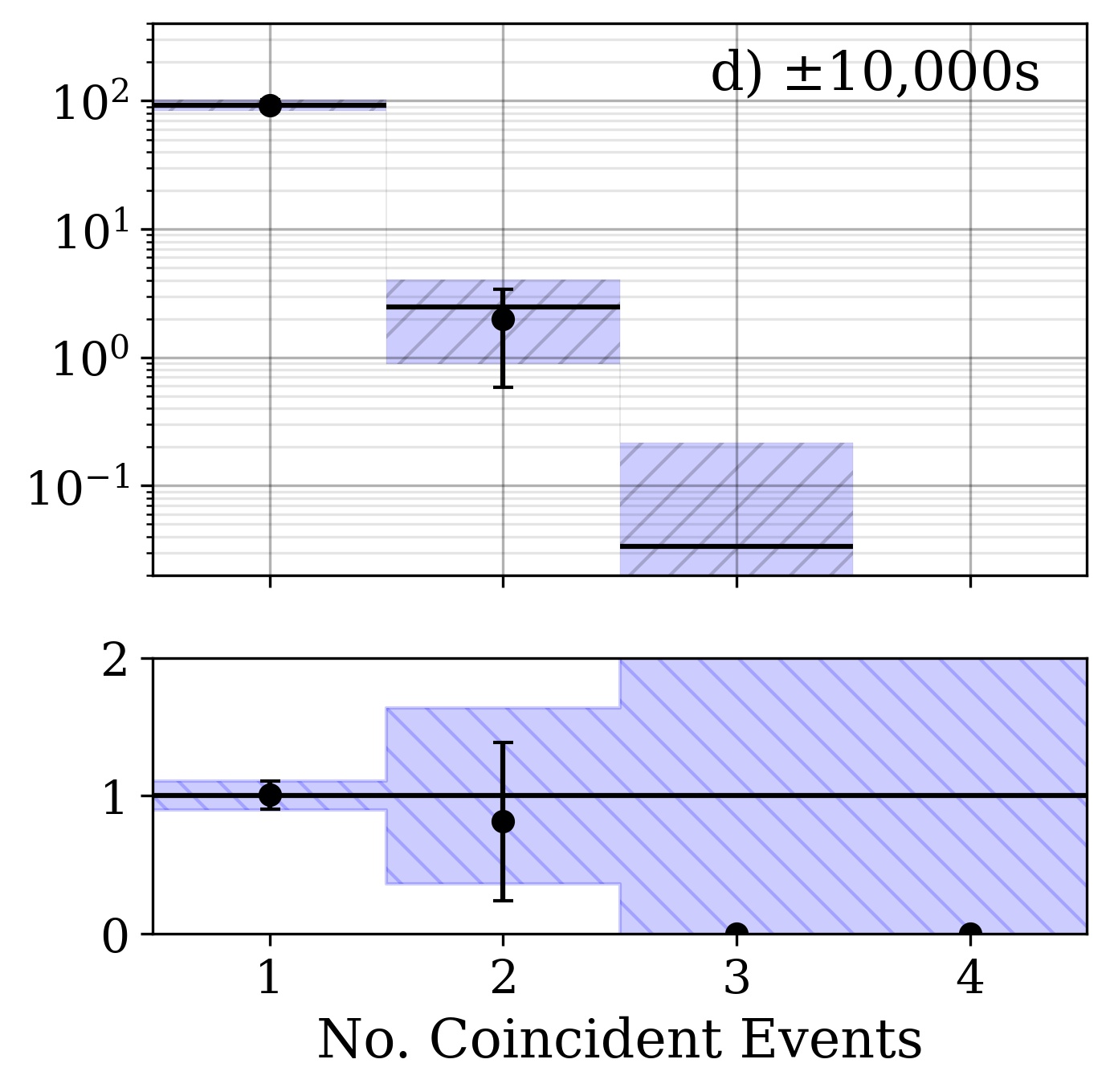}
\end{subfigure}
\caption{The number of observed signal candidates are shown as black data points with statistical uncertainties for each of the four predefined time windows: (a)\,$\pm$500s, (b)\,$\pm$1,000\,s, (c)\,$\pm$3,600\,s, and (d)\,$\pm$10,000\,s. The black horizontal lines represent the expected number of events given the Poissonian background from each period reported in Table\,\ref{table::PeriodData}, along with the hatched area representing the statistical uncertainty. The ratio of the number of observed signal candidates to the background expectation is shown in the lower subplots.}
\label{fig::time}
\end{figure}

\section{Results}


The number of observed signal candidates in the predefined time windows, compared to the Poisson-fluctuated IBD background expectation, is shown in Fig.\,\ref{fig::time}. No significant observation above the expected background is observed in any time window. Zero coincident events were observed using time windows (a) and (b). A single IBD event was observed in coincidence with a high-energy IceCube event (IceCube run number\,132229 and event number\,66688965) in time window (c). Time window (d) observed a second coincident event (IceCube run number\,134817 and event number\,29175858). These observations, however, are compatible with the background expectation for time windows (c) and (d): 0.94 and 2.51 events, respectively. The first coincident event occurred on 21\,February\,2019\,(Period IV), at 07:51:55.569\,(UTC). The IceCube track-like neutrino\footnote{\url{https://gcn.gsfc.nasa.gov/gcn3/23918.gcn3}} was categorized as a Gold event and reported to have an energy of approximately 56\,TeV. The KamLAND IBD interaction occurred 2024\,s prior to the IceCube neutrino and had an energy of E$_{\bar{\nu}_e} = 5.2$\,$\pm$\,$0.1$\,MeV. The second coincident event occurred on 21\,December\,2020 (Period IV) at 15:16:24.736\,(UTC). The IceCube event was also track-like, \footnote{\url{https://gcn.gsfc.nasa.gov/gcn3/29102.gcn3}} categorized as a Gold event, and had a measured energy of approximately 175\,TeV. The KamLAND neutrino had a reconstructed energy of E$_{\bar{\nu}_e} = 3.1$\,$\pm$\,$0.1$\,MeV and arrived 9571.3\,s after the IceCube neutrino.


The closest IBD timestamp to IceCube-170922A arrived 222,048\,s (2.57\,days) prior to the IceCube event. The Poisson probability of observing a single coincident event arriving within this time window is approximately 23\%. 
The high-energy gamma radiation observed by MAGIC (E$_\gamma>$\,90\,GeV), H.E.S.S.\,(E$_\gamma>$\,175\,GeV), and VERITAS\,(E$_\gamma>$\,175\,GeV) found that the TXS\,0506+056 blazar was in a high-emission state for $\sim$12 days after the observation of IceCube-170922A. In this time span, KamLAND observed two IBD events occurring 8.73 and 9.35 days after IceCube-170922A. These two events had an energy of $E_{\bar{\nu}_e} = 2.4$\,$\pm$\,$0.1$ and $3.1$\,$\pm$\,$0.1$, respectively. This observation is also consistent with the IBD background expectation, and the Poisson probability of observing two events within this time window is 21\%.



\begin{figure}[h]
\centering
\begin{subfigure}
  \centering
  \includegraphics[width=0.463\linewidth]{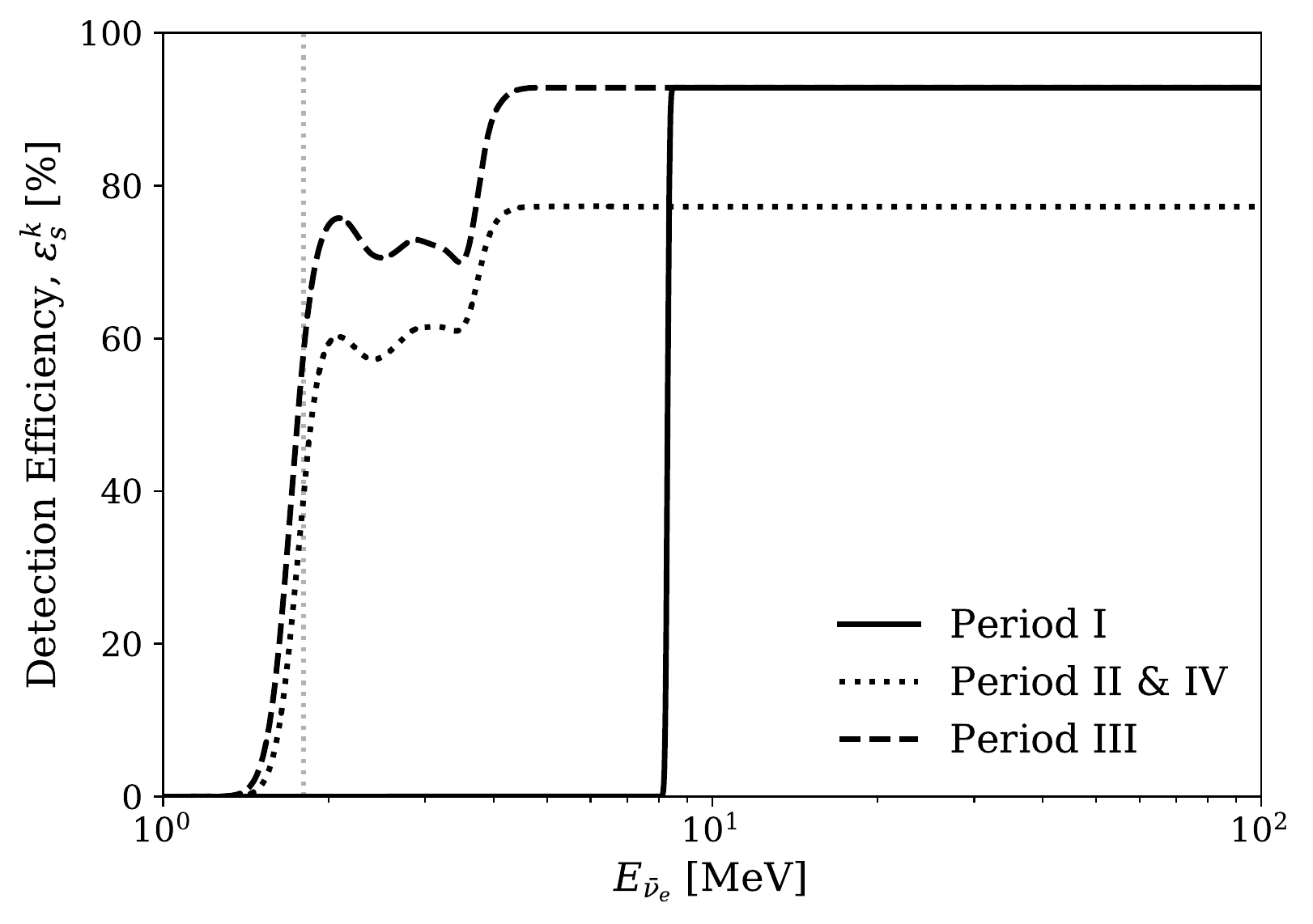}
\end{subfigure}
\begin{subfigure}
  \centering
  \includegraphics[width=.48\linewidth]{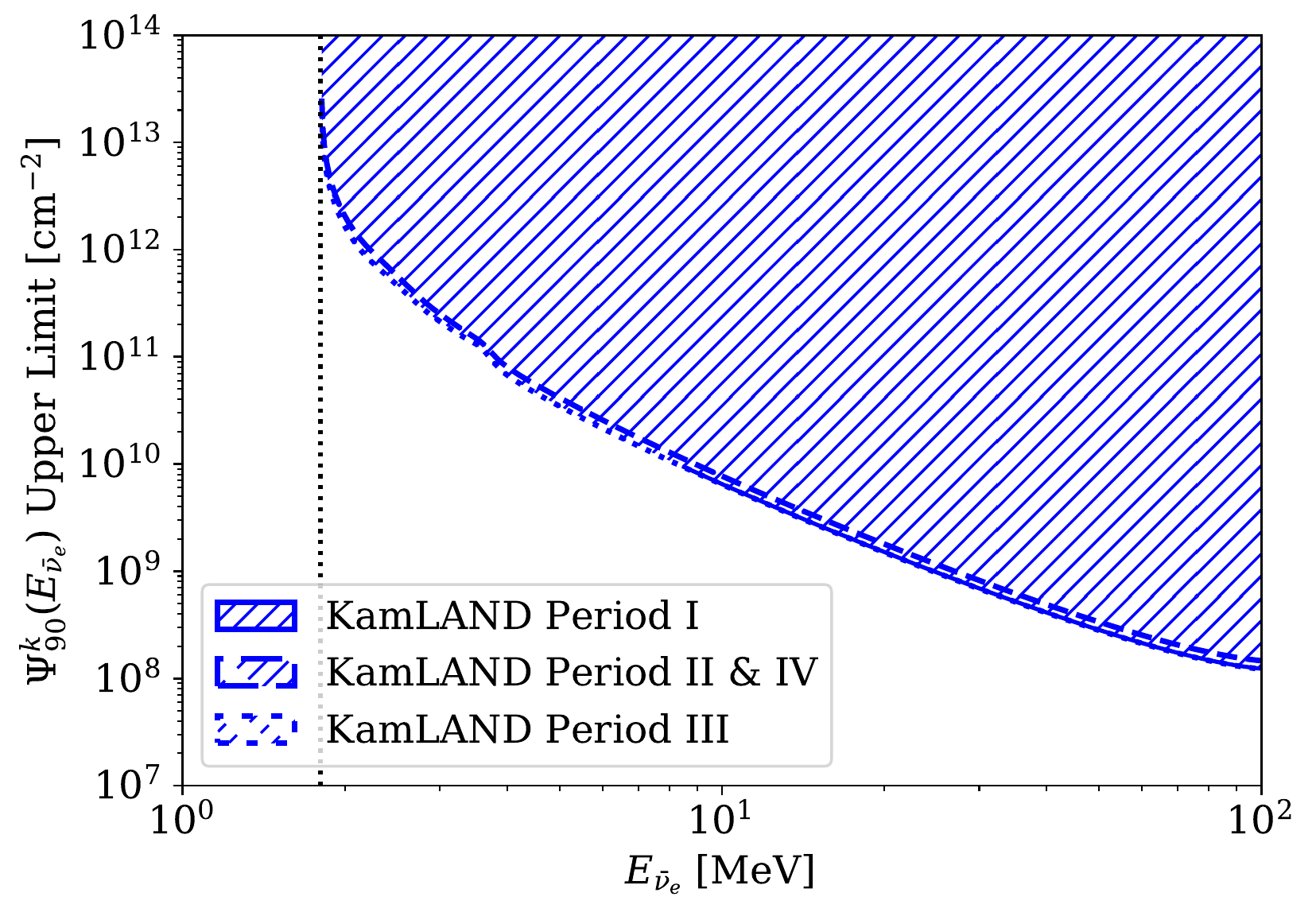}
\end{subfigure}
\caption{Left: The IBD detection efficiencies as a function of reconstructed E$_{\bar{\nu}_e}$ for each KamLAND period. 
Above  5\,MeV, the detection efficiencies converge to approximately 92.9\% and 77.4\% for Periods I\,\&\,III and II\,\&\,IV, respectively and are constant up to 100\,MeV. The structure below 4\,MeV is primarily due to the likelihood-based selection criteria. The non-zero detection efficiency below the low-energy IBD threshold arises from the finite energy resolution in KamLAND. Right: The  90\% C.L. upper limit on the mono-energetic $\bar{\nu}_e$ fluence, for energies between $E_{\bar{\nu}_e} = 1.8$\,MeV and 100\,MeV, and for t$_p=500$s. The results of Periods II\,\&\,IV are within the line width of each other. In both figures, a vertical dotted line is shown at the IBD threshold.}
\label{fig::fluence}
\end{figure}

In the absence of a signal, we present the $90\%$ C.L. upper limit on the number of observed signal neutrinos, $N_{90}$, for the $\pm$500\,s time window using the Feldman-Cousins method \cite{PhysRevD.57.3873}. The period-dependent values are listed in the right-most column of Table\,\ref{table::PeriodData}. The 90\% C.L. upper limit on the $\bar{\nu}_e$ fluence for each high-energy IceCube neutrino in period $k$ is then given by:
\begin{equation}\label{eq::fluence}
F_{90}^k = \frac{N_{90}^k} {N_T  \int_{E_l}^{E_h}  \sigma(E_{\bar{\nu}_e}) \lambda(E_{\bar{\nu}_e})  \epsilon^k_s(E_{\bar{\nu}_e}) dE_{\bar{\nu}_e} },
\end{equation}
where the integral is performed over the energy range of this analysis ($E_l = 1.8$\,MeV and $E_h = 100$\,MeV), $N_{T}$ = 5.98$\times10^{31}$ is the number of target protons in the fiducial volume, $\sigma$($E_{\bar{\nu}_e}$) is the IBD cross-section~\cite{strumia2003precise}, $\lambda (E_{\bar{\nu}_e})$ is the normalized $\bar{\nu}_e$ energy spectrum, and $\epsilon^k_s$ is the energy-dependent IBD detection efficiency found in Fig.~\ref{fig::fluence}\,(left). Prior to making an assumption on $\lambda(E_{\bar{\nu}_e})$, we compute the equivalent fluence Green's function using a mono-energetic neutrino flux by replacing $\lambda(E_{\bar{\nu}_e})$ with a delta function:
\begin{equation}\label{eq::greenes}
\Psi_{90}^k(E_{\bar{\nu}_e}) = \frac{N_{90}^k} {N_T  \int_{E_l}^{E_h}   \sigma(E^{'}_{\bar{\nu}_e}) \delta(E_{\bar{\nu}_e}-E^{'}_{\bar{\nu}_e})  \epsilon^k_s(E^{'}_{\bar{\nu}_e}) dE^{'}_{\bar{\nu}_e}}.
\end{equation}
Fig.\,\ref{fig::fluence}\,(right) shows the 90\% C.L. upper limit on the mono-energetic $\bar{\nu}_e$ fluence, for each of the four KamLAND periods. The curves shown here are also a close approximation of the experimental sensitivity for each period. We note that had we observed a single coincident event with time window (a), the significance of this observation would exclude the null hypothesis at the 99\% C.L. 
The presented limits modestly change when using the more conservative time windows. For Periods I\,-\,III the limits are found to become more stringent by less than 10\% for all time windows. The observation of a single coincident event using time window (c) and two events using time window (d), increase the Period IV limit by 59\% and 82\%, respectively.

\begin{figure}[t]
\centering
\begin{subfigure}
  \centering
  \includegraphics[width=0.48\linewidth]{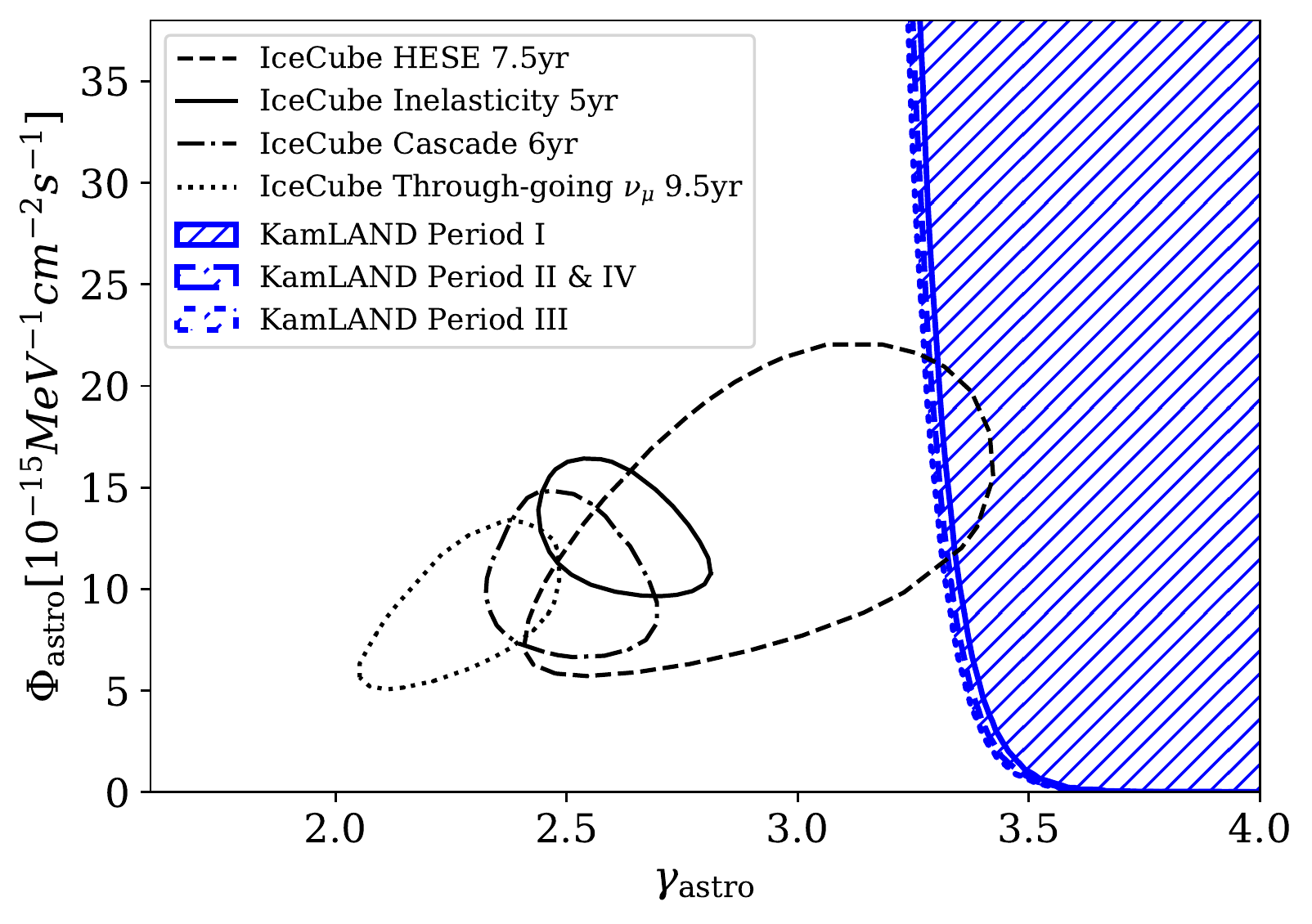}
\end{subfigure}
\begin{subfigure}
  \centering
  \includegraphics[width=.48\linewidth]{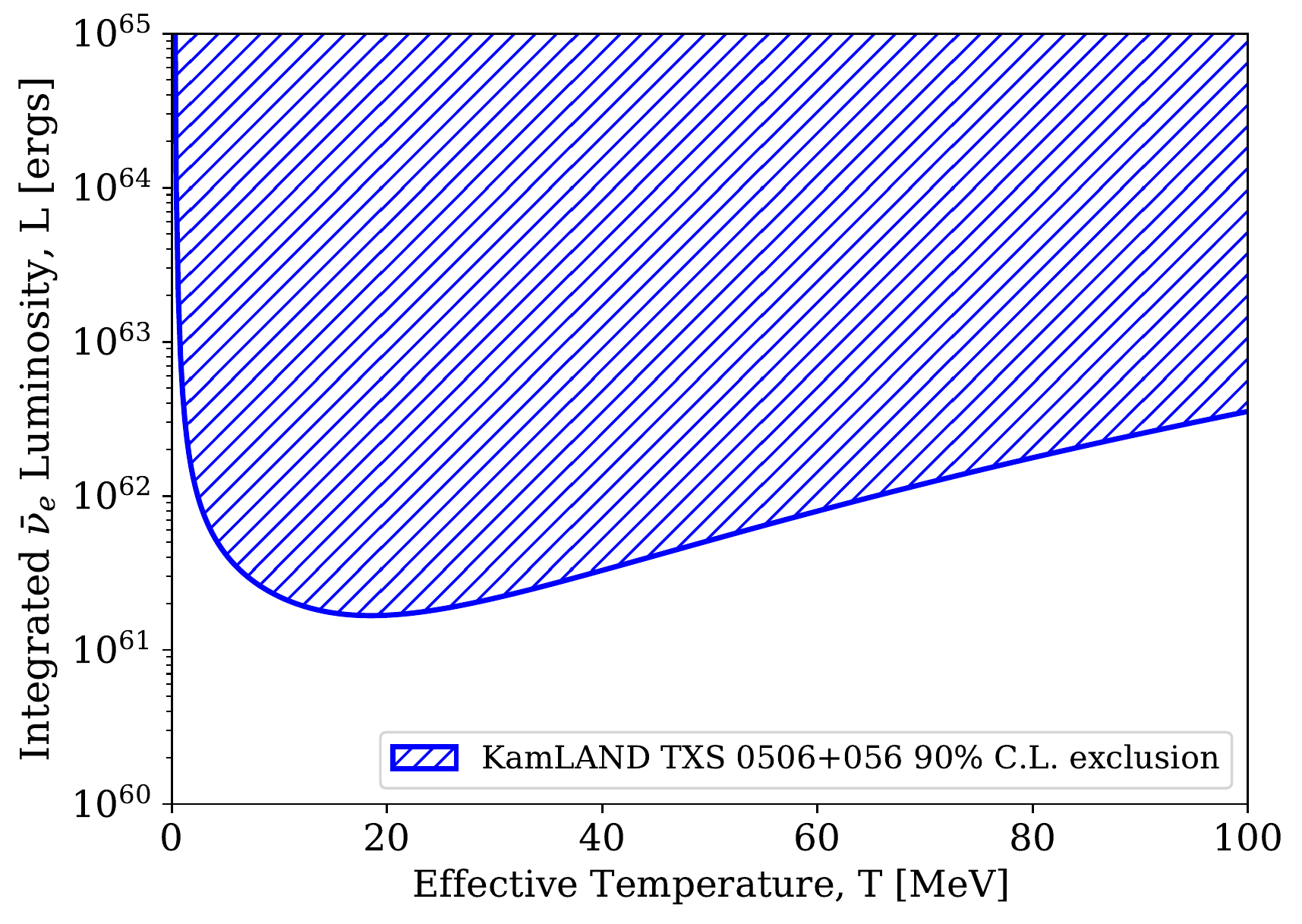}
\end{subfigure}
\caption{Left: The excluded astrophysical flux parameters, $\Phi_{\mathrm{astro}}$ and $\gamma_{\mathrm{astro}}$, at the 90\% C.L. from KamLAND, assuming an unbroken power-law energy spectrum. Also shown are the IceCube's 95.4\% confidence regions from the 7.5yr HESE sample~\cite{abbasi2021icecube}, 
the 5yr inelasticity measurement~\cite{aartsen2019measurements}, the 6yr cascade sample~\cite{aartsen2020characteristics}, and the 9.5yr Northern track sample preliminary results~\cite{stettner2019measurement}. The normalization constant is presented per neutrino species, integrated over the full sky. Right: The 90\% C.L. excluded $\bar{\nu}_e$ luminosity and effective source temperature, assuming a Fermi-Dirac energy distribution at the distance of the TXS 0506+056 blazar.}
\label{fig::astro}
\end{figure}

The neutrino flux is often modeled as an isotropic unbroken power-law spectrum~\cite{aartsen2016observation} defined by a normalization constant at 100\,TeV, $\Phi_{\mathrm{astro}}$, and spectral index, $\gamma_{\mathrm{astro}}$:
\begin{equation}\label{eq::PL}
\lambda_{PL}(E_{\bar{\nu}_e},\Phi_{\mathrm{astro}}, \gamma_{\mathrm{astro}} ) = \Phi_{\mathrm{astro}} \Big( \frac{E_{\bar{\nu}_e}}{100\mathrm{TeV}}\Big)^{-\gamma_{\mathrm{astro}}}.
\end{equation}
This is the model that IceCube primarily employs to fit their high-energy neutrino datasets~\cite{abbasi2021icecube,aartsen2019measurements,aartsen2020characteristics,stettner2019measurement}. We can extrapolate our null observation limits to higher energies to compare with the IceCube results. Using the derived $90\%$ C.L. upper limit on the number of observed signal neutrinos in period $k$, the excluded region in terms of the parameter space described in Eq.~\ref{eq::PL}, is calculated as:
\begin{equation}\label{eq::lum}
N_{90}^{k} = N_T \int_{E_l}^{E_h}  \epsilon^k_s(E_{\bar{\nu}_e}) \sigma(E_{\bar{\nu}_e}) \lambda(E_{\bar{\nu}_e}) dE_{\bar{\nu}_e},
\end{equation}
where we set $\lambda(E_{\bar{\nu}_e}) = \lambda_{PL}(E_{\bar{\nu}_e},\Phi_{\mathrm{astro}}, \gamma_{\mathrm{astro}})$. Fig.~\ref{fig::astro} (left) shows the excluded part of the unbroken power-law spectrum for a single astrophysical neutrino, compared to the allowed 95.4\% confidence region per neutrino species from IceCube. The slight tension between the different IceCube measurements could be an indication that a simple unbroken power-law fit is not sufficient to describe the astrophysical neutrino flux spectrum. As in the case of the astrophysical cosmic rays, it could be the case that there is a break or low-energy cut-off in the spectrum below the high-energy domain of IceCube and the low energies observed in KamLAND.

The high-temperature environment resulting from the rapidly accreting material surrounding
an AGN may emit MeV-scale thermal neutrinos. The thermal emission from other sources of accretion disks (i.e. collapsars~\cite{nagataki2002features}, supernovae~\cite{sekiguchi2011formation, mclaughlin2007supernova}, and binary mergers~\cite{caballero2009detecting}) are often modeled as a Fermi-Dirac distribution~\cite{abbasi2021search,kotake2006explosion,caballero2009detecting} at an effective temperature, $T$ and zero chemical potential:
\begin{equation}
\Phi_{FD}(E_{\bar{\nu}_e},T) =  \frac{1}{T^3 f_2}\frac{E_{\bar{\nu}_e}^2}{e^{(E_{\bar{\nu}_e}/T)}+1}, \quad f_n = \int_0^{\infty} \frac{x^n}{e^{x} +1} dx.
\end{equation}
The at-Earth flux, given the redshift and luminosity distance to TXS 0506+056, is calculated through: 
\begin{equation}
\lambda_{FD}(E_{\bar{\nu}_e},T,L) =\frac{1 + z}{4\pi d^4} \frac{L}{\langle{}E\rangle{}} \Phi_{FD}((1+z)E_{\bar{\nu}_e},T),
\end{equation}
where $L$ is the isotropic source $\bar{\nu}_e$ luminosity. As in Ref.~\cite{asakura2015study}, we relate the mean $\bar{\nu}_e$ energy to the effective temperature through $\langle{}E\rangle{} = 3.15 T$. We now set $\lambda(E_{\nu}) = \lambda_{FD}(E_{\bar{\nu}_e},T,L)$ in Eq.~\ref{eq::lum} to compute the 90\%~C.L. limits on the integrated $\bar{\nu}_e$ luminosity as a function of source effective temperature. Fig.~\ref{fig::astro} (right) shows the excluded region for TXS 0506+056 based on the null observation. The presented limit assumes an isotropic thermal neutrino emission. Due to the large distance to the TXS 0506+056 blazar, the limits are approximately nine orders of magnitude higher than the $\bar{\nu}_e$ luminosity/temperature of supernova SN 1987A~\cite{hirata1988observation, schramm1987neutrinos, lattimer1989analysis}. 

\section{Conclusion} \label{sec:conclusion}
In this analysis, we performed a search for low-energy electron antineutrinos in KamLAND correlated with the publicly available high-energy neutrino datasets from the IceCube Neutrino Observatory. The analysis examined 102 high-energy neutrino events in IceCube for correlations with 341 low-energy KamLAND neutrinos, spanning from October 2010 to June 2021. No significant excess above the expected background was observed using a coincident time window of  $\pm$500s, $\pm$1,000\,s, $\pm$3,600\,s and  $\pm$10,000\,s. Given the null observation, the 90\% C.L. upper limit  assuming a mono-energetic neutrino flux was presented. A comparison to the measured IceCube astrophysical flux assuming an unbroken power-law energy spectrum was also performed. Finally, using the redshift to the TXS 0506+056 blazar, we also presented limits on the isotropic thermal MeV-scale neutrino emission assuming a Fermi-Dirac energy spectrum.

\section{Acknowledgments}

The KamLAND experiment is supported by JSPS KAKENHI Grants 19H05803; the World Premier International Research Center Initiative (WPI Initiative), MEXT, Japan; Dutch Research Council 
(NWO); and under the U.S. Department of Energy (DOE) Contract No.~DE-AC02-05CH11231, the National Science Foundation (NSF) No.~NSF-1806440,~NSF-2012964, the Heising-Simons Foundation, as well as other DOE and NSF grants to individual institutions. The Kamioka Mining and Smelting Company has provided services for activities in the mine. We acknowledge the support of NII for SINET4. 

\appendix

 \bibliographystyle{elsarticle-num} 
 \bibliography{cas-refs}





\end{document}